\vsize=700 pt
\headline={\ifnum\pageno>1 \hss \number\pageno\ \hss \else\hfill \fi}
\pageno=1
\nopagenumbers
%----------------------------------------------------------------------------
\centerline{ \bf THE COMPLETE COHOMOLOGY OF $E_8$ LIE ALGEBRA}
\vskip 15mm
\centerline{\bf H. R. Karadayi and M. Gungormez}
\centerline{Dept.Physics, Fac. Science, Tech.Univ.Istanbul }
\centerline{ 80626, Maslak, Istanbul, Turkey }
\centerline{ Internet: karadayi@sariyer.cc.itu.edu.tr}
\vskip 10mm
\centerline{\bf{Abstract}}
\vskip 10mm

It is shown, for any irreducible representation of $E_8$ Lie algebra,
that eigenvalues of Casimir operators can be calculated in the form of
invariant polinomials which are decomposed in terms of $A_8$ basis functions.
The general method is applied for degrees 8,12 and 14 for which  2,8 and 19
invariant polinomials are obtained respectively. For each particular degree,
these invariant polinomials can be taken to be $E_8$ basis functions in the
sense that any Casimir operator of $E_8$ has always eigenvalues which can
be expressed as linear superpositions of them. This can be investigated
by showing that each one of these $E_8$ basis functions gives us a linear
equation to calculate weight multiplicities.

\vskip 15mm
\vskip 15mm
\vskip 15mm
\vskip 15mm
\vskip 15mm
\vskip 15mm
\vskip 15mm
\vskip 15mm
\vskip 15mm
\vskip 15mm

\hfill\eject

\vskip 3mm
\noindent {\bf{I.\ INTRODUCTION}}
\vskip 3mm
The cohomology for Lie algebras states non-linear relationships between
elements of the center of their universal enveloping algebras {\bf [1]}.
The non-linearity comes from the fact that these relationships are between
the elements of different orders and the non-linearly independent ones are
determined by their Betti numbers {\bf [2]}. A problem here is to determine
the number of linearly independent elements of the same order. In previous
works {\bf [3]}, we studied this problem for $A_N$ Lie algebras and give a
method to obtain, in any order and for any irreducible representation, the
eigenvalues of Casimir operators. This will be extended here to $E_8$ Lie
algebra.

$E_8$ is the biggest one of the finite dimensional Lie algebras and plays
a striking role in high energy physics as well as in mathematics. It provides
a natural laboratory to study the structure of $E_{10}$ hyperbolic Lie
algebra {\bf [4]} which is seen to play a key role in understanding the
structure of infinite dimensional Lie algebras beyond affine Kac-Moody Lie
algebras. There are so much works to show the fact that  its significance
in string theories and in the duality properties of supersymmetric gauge
theories is now well-established. An explicit calculation of the eigenvalues
of $E_8$ Casimir operators is therefore worthwhile to study.

It is known that, beyond degree 2, $E_8$ Betti numbers give us non-linearly
independent Casimir elements for degrees  8,12,14,18,20,24,30. The method
given here can be applied equally well for all these degrees. Our results
will, however, be presented only for degrees 8,12 and 14. To this end,
we obtain respectively 2, 8 and 19 different invariant polinomials which
say us that there are 2, 8 and 19 numbers of linearly independent Casimir
elements of orders 8, 12 and 14. A remark here is that these polinomials are
expressed by decomposing them in terms of some $A_8$ basis functions. The
very meaning of this is as in the following: such a decomposition is valid
for any irreducible $E_8$ representation but {\bf its coefficients does not
yield a dependence on the representations}. This non-trivial property
specifies, in fact, our definition of $A_8$  basis functions.

A technical aside of our work is to find a way of classifying the weights
participating within a Weyl orbit. This will be given here in a manner which
is convenient for our purposes though there are some efforts {\bf [5]} in the
litterature concerning this point. In terms of $A_8$ subalgebra, weight
classifications of $E_8$ Weyl orbits are to be carried out by permutations
just as in $A_N$ Lie algebras. This will be given in section (II) in the
form of a second permutational lemma. In section (III), the procedure which
we previously develop {\bf [3]} for $A_N$ Lie algebras, will be applied here
to obtain the so-called Weyl orbit characters of $E_8$ Lie algebra.
In section (IV), we show that $E_8$ basis functions can be decomposed in
terms of some properly chosen $A_8$ basis functions. Some conclusions are
given in the last section. The detailed expressions are to be given in
the 4 appendices.

\vskip 3mm
\noindent {\bf{II.\ WEIGHT CLASSIFICATION OF $E_8$ WEYL ORBITS}}
\vskip 3mm
We refer the excellent book of Humphreys {\bf[6]} for technical aspects of
this section. A brief summary of our framework will also be given here though
a comprehensive account of subject is given in our previous work {\bf [3]}.
It is known that the weights of an irreducible representation $R(\Lambda^+)$
can be decomposed in the form of
$$ R(\Lambda^+) = \Pi(\Lambda^+) \ \ + \sum m(\lambda^+ < \Lambda^+) \ \
\Pi(\lambda^+) \eqno(II.1) $$
where $\Lambda^+$ is the principal dominant weight of the representation,
$\lambda^+$'s are their sub-dominant weights and $m(\lambda^+ < \Lambda^+)$'s
are multiplicities of weights $\lambda^+$ within the representation
$ R(\Lambda^+) $. Once a convenient definition of eigenvalues is assigned
to $ \Pi(\lambda^+) $, it is clear that this also means for the whole
$R(\Lambda^+)$ via (II.1).

In the conventional formulation, it is natural to define Casimir eigenvalues
for irreducible representations which are known to have matrix representations.
A known exception is the second order Casimir invariant which are defined
directly by its own eigenvalues. It is introduced by Kac {\bf[7]} to expose
the structure theory of representations of Kac-Moody Lie algebras. In favor
of the existence of a permutational lemma, the extension of eigenvalue
concept can be made for $A_N$ Weyl orbits. This however could not be so
clear for Lie algebras other than $A_N$ Lie algebras. To this end, we will
now give a second permutational lemma. To be more concrete, we proceed
in terms of $A_8$ decomposition of $E_8$ Weyl orbits. Our framework is
to obtain a covering
$$ \Pi(\lambda^+) \equiv  \sum_{\sigma^+ \in \Sigma(\lambda^+)} \Pi(\sigma^+)
\eqno(II.2) . $$
where $\Sigma(\lambda^+)$ is the set of all $A_8$ dominant weights
participating within the same $E_8$ Weyl orbit $\Pi(\lambda^+)$. If one is
able to determine the set $\Sigma(\lambda^+)$ completely, the weights of each
particular $A_8$ Weyl orbit $\Pi(\sigma)$ and hence the whole $\Pi(\lambda^+)$
are known. We thus extend the eigenvalue concept to $E_8$ Weyl orbits just as
in the case of $A_N$ Lie algebras.

It is known, on the other hand, that the elements of $\Sigma(\lambda^+)$ have
the same square length with $E_8$ dominant $\lambda^+$. It is unfortunate
that this sole condition remains unsufficient to obtain the whole structure
of the set $\Sigma(\lambda^+)$. This exposes more severe problems especially
for Lie algebras having Dynkin diagrams with higher degree of diagram
automorphisms, for instance Kac-Moody algebras. To solve this non-trivial
part of the problem, we introduce 9 fundamental weights $\mu_I$ of $A_8$,
via scalar products
$$ \kappa(\mu_I,\mu_J) \equiv \delta_{IJ} - {1 \over 9} \ \ , \ \
I,J = 1,2, .. 9 \eqno(II.3)  $$
The existence of $\kappa(.,.)$ is known to be guaranteed by $A_8$ Cartan matrix.
The fundamental dominant weights of $A_8$ are now expressed by
$$  \sigma_i \equiv \sum_{j=1}^{i} \mu_j \ \ , \ \ i=1,2, .. 8. \eqno(II.4)  $$
and the correspondence between $E_8$ and $A_8$ Lie algebras is provided by
$$ \eqalign{
\lambda_1 &= \sigma_1 + \sigma_8 \cr
\lambda_2 &= \sigma_2 + 2 \ \sigma_8  \cr
\lambda_3 &= \sigma_3 + 3 \ \sigma_8  \cr
\lambda_4 &= \sigma_4 + 4 \ \sigma_8  \cr
\lambda_5 &= \sigma_5 + 5 \ \sigma_8  \cr
\lambda_6 &= \sigma_6 + 3 \ \sigma_8  \cr
\lambda_7 &= \sigma_7 + \sigma_8  \cr
\lambda_8 &= 3 \ \sigma_8   }  \eqno(II.5)    $$
where $\lambda_i$'s are the fundamental dominant weights of $E_8$ for which
any dominant weight has thus the form
$$ \Lambda^+ \equiv \sum_{i=1}^8 r_i \lambda_i  \ \ , \ \ r_i \in Z^+ .
\eqno(II.6) $$
$Z^+$ here is the set of positive integers including zero. As will be seen in
the following lemma, the primary knowledge we need here is only the explicit
tabulation of elements of the {\bf fundamental sets} $\Sigma(\lambda_i)$
for i=1,2, .. 8. For instance,
$$ \Sigma(\lambda_1) = ( \sigma_1 + \sigma_8 \ , \ \sigma_3 \ , \ \sigma_8 ) $$
for which we have the decomposition
$$ \Pi(\lambda_1) = \Pi(\sigma_1+\sigma_8) + \Pi(\sigma_3) + \Pi(\sigma_6)
\eqno(II.7) $$
of 240 roots of $E_8$ Lie algebra. Due to permutational lemma in ref(3),
$A_8$ Weyl orbits here are known to have the weight structures
$$ \eqalign{
\Pi(\sigma_1+\sigma_8) &= (\mu_{I_1}+\mu_{I_2}+\mu_{I_3}+\mu_{I_4}+
\mu_{I_5}+\mu_{I_6}+\mu_{I_7}+\mu_{I_8}) \cr
\Pi(\sigma_3) &= (\mu_{I_1}+\mu_{I_2}+\mu_{I_3})  \cr
\Pi(\sigma_6) &= (\mu_{I_1}+\mu_{I_2}+\mu_{I_3}+\mu_{I_4}+\mu_{I_5}+\mu_{I_6}) }
\eqno(II.8) $$
where all indices are permutated over the set (1,2, .. 9) providing no two of
them are equal. Note here by (II.4) that
$$ \eqalign{
\sigma_1+\sigma_8 &= \mu_1+\mu_2+\mu_3+\mu_4+\mu_5+\mu_6+\mu_7+\mu_8 \cr
\sigma_3 &= \mu_1+\mu_2+\mu_3  \cr
\sigma_6 &= \mu_1+\mu_2+\mu_3+\mu_4+\mu_5+\mu_6 }  \eqno(II.9)\ . $$
The formal similarity between (II.8) and (II.9) is a resume of the
first permutational lemma.

\hfill\eject

\noindent Let us now extend it to $E_8$ Lie algebra:
\par {\bf For any dominant weight having the form (II.6), the set
$\Sigma(\Lambda^+)$ of $A_8$ dominant weights is
\par specified by
$$ \Sigma(\Lambda^+) = \sum_{i=1}^8 r_i \ \Sigma(\lambda_i). \eqno(II.10) $$
\par together with the equality of square lengths.}

\noindent In addition to $\Sigma(\lambda_1)$ given above, the other
7 fundamental sets $\Sigma(\lambda_i)$ have respectively 7,15,27,35,17,5
and 11 elements for i=2,3, .. 8 as they will be given in appendix(1).
It is therefore clear that the weight decomposition of any $E_8$ Weyl orbit
is now completely known in terms of $A_8$ Weyl orbits in the presence of both
of our lemmas.

\vskip 3mm
\noindent {\bf{III.CALCULATION OF $E_8$ ORBIT CHARACTERS}}
\vskip 3mm
It is now crucial for practical use to express an $A_8$ dominant
weight $\sigma^+$ in the form
$$ \sigma^+ \equiv \sum_{i=1}^8 q_i \ \mu_i \ \ \ , \ \ \
q_1 \geq q_2 \geq ... \geq q_8 \geq 0  \ \   . \eqno(III.1) $$
To prevent repetitions, we only give here the main definitions and formulas
which can be obtained from those in ref(3) for $A_8$. It is then known for
a Weyl orbit $\Pi(\sigma^+)$ that eigenvalues of a Casimir operator of
degree M are to be defined by the aid of the formal definition
$$ ch_M(\sigma^+) \equiv \sum_{\mu \in \Pi(\sigma^+)} (\mu)^M \eqno(III.2) $$
where the right hand side of (III.2) is to be determined by its decomposition
in terms of some generators given in appendix(2). These generators can
be reduced in terms of the following ones which are of main interest for our
purposes:
$$ \eqalign{
Q(M) &\equiv \sum_{i=1}^8 (q_i)^M \cr
\mu(M) &\equiv \sum_{I=1}^9 (\mu_I)^M  } \ \ . \eqno(III.3) $$
We remark here by definition that $\mu(1) \equiv 0 $ and hence
$$ (\mu(1))^M \equiv 0 \ \ , \ \ M=1,2, .. ,9,10, ... . \eqno(III.4) $$
It can be readily seen that (III.4) are to be fulfilled automatically for
M=2,3, .. 9 but gives rise to the fact that all the generators $\mu(M)$ for
$M \geq 10$ are non-linearly depend on the ones for M=2,3, .. 9. These
non-linearities are clearly the reminiscents of $A_8$ cohomology. We call
them {\bf $A_8$ Dualities} because (III.4)  allows the existence of a
Levi-Civita tensor in 9 dimensions. The first example is
$$ \eqalign{
\mu(10) &\equiv {1  \over 8!} \ (25200 \ \mu(2)   \ \mu(8)
+ 19200 \ \mu(3)   \ \mu(7)
+ 16800 \ \mu(4)   \ \mu(6)                               \cr
&- 8400  \ \mu(2)^2 \ \mu(6)
- 13440 \ \mu(2)   \ \mu(3) \ \mu(5)
+ 8064  \ \mu(5)^2
+ 2100  \ \mu(2)^3 \ \mu(4)                               \cr
&- 5600  \ \mu(3)^2 \ \mu(4)
- 6300  \ \mu(2)   \ \mu(4)^2 + 2800  \ \mu(2)^2 \ \mu(3)^2
- 105   \ \mu(2)^5\ )  }  \eqno(III.5) $$
It is seen that $\mu(10)$ consists of p(10)=11 monomials coming from the
par\-ti\-ti\-ons of 10 into the set \hfill\break (2,3,4,5,6,7,8,9) and the other ones can be
calculated as in the following:
$$ \eqalign{ p(8) &= 7 \ , \ p(10) = 11 \ , \ p(12) = 19 \ , \ p(14) = 29 \ , \
p(16) = 44 \ , \ p(18) = 66  \cr
p(20) &= 94 \ , \ p(22) = 131 \ , \ p(24) = 181 \ , \ p(26) = 239 \ , \
p(28) = 309 \ , \ p(30) = 390    }   \eqno(III.6)   $$
and
$$ \eqalign{ p(11) &= 13 \ , \ p(13) = 21 \ , \ p(15) = 34 \ , \
p(17) = 51 \ , \ p(19) = 75   \cr
p(21) &= 109 \ , \ p(23) = 151 \ , \ p(25) = 204 \ ,
\ p(27) = 270 \ , \ p(29) = 344  } \eqno(III.7)   $$
Within the scope of this work, we also need the generators $\mu(11) \ , \
\mu(12) \ , \ \mu(14)$. They are given in appendix(3).

\vskip 3mm
\noindent {\bf{IV.DECOMPOSITIONS OF INVARIANT POLINOMIALS IN THE $A_8$ BASIS}}
\vskip 3mm
Let us begin with the example of degree 8 for which (III.2)  has the decomposition
$$ ch_8(\Lambda^+) \equiv \sum_{\alpha=1}^7 cof_\alpha(\Lambda^+) \  T(\alpha)
\eqno(IV.1) $$
where 7 generators $T(\alpha)$ come from the 7 monomials mentioned above.
The invariant polinomials $P_\alpha(\Lambda^+)$ which give us
Casimir eigenvalues stem from the coefficients $cof_\alpha(\Lambda^+)$
via definitions
$$ P_\alpha(\Lambda^+) \equiv
{cof_\alpha(\Lambda^+) \over cof_\alpha(\lambda_1)} \
{dimR(\lambda_1) \over dimR(\Lambda^+)} \
P_\alpha(\lambda_1)   \eqno(IV.2) $$
where $dimR(\Lambda^+)$ is the dimension of the representation $R(\Lambda^+)$.
An important notice here is the fact that we do not need here the Weyl
dimension formula. This will be provided by orbital decomposition (II.1)
providing the sets $\Sigma(\lambda^+)$ are known for each particular
subdominant
$\lambda^+$ of $\Lambda^+$. Let us recall from ref(3) that we can
calculate by permutations the dimensions of $A_8$ Weyl orbits.
The fundamental representation $R(\lambda_1)$ of $E_8$ is taken in (IV.2) to be
a reference representation, i.e all our expressions for Casimir eigenvalues
are to be given by normalizing with respect to fundamental representation.
Explicit calculations now show that we can find only 2 different polinomials
among 7 invariant polinomials $P_\alpha(\Lambda^+)$:
$$ {\cal K}_1(8, \Lambda^+) \equiv 729 \ {\it \Theta}(8, \Lambda^+) - 71757069294212 \eqno(IV.3)  $$
from the monomial $\mu(2)^4$ and
$$ \eqalign{
{\cal K}_2(8, \Lambda^+) &\equiv  68580 \ {\it \Theta}(8, \Lambda^+) \ - \cr
&42672 \ {\it \Theta}(6, \Lambda^+) \ {\it \Theta}(2, \Lambda^+)  \ - \cr
&42672 \ {\it \Theta}(5, \Lambda^+) \ {\it \Theta}(3, \Lambda^+)  \ -  \cr
&13335 \ {\it \Theta}(4, \Lambda^+)^2       \ +             \cr
&13335 \ {\it \Theta}(4, \Lambda^+) \ {\it \Theta}(2, \Lambda^+)^2  \ +  \cr
&17780 \ {\it \Theta}(3, \Lambda^+)^2 \ {\it \Theta}(2, \Lambda^+)  \ -   \cr
&939 \ {\it \Theta}(2, \Lambda^+)^4       \ +   \cr
&385526887200     }  \eqno(IV.4)  $$
which is common for all other monomials.

The functions $ {\it \Theta}(M, \Lambda^+)$ are {\bf $A_8$ basis functions} which are
defined by
$$ {\it \Theta}(M, \Lambda^+) \equiv \sum_{I=1}^9 ( \theta_I(\Lambda^+))^M  \ \ \ ,
\ \ \ M = 1,2, ...  \eqno(IV.5)   $$
where
$$ \eqalign{
\theta_1(\Lambda^+) &\equiv
{19 \over 3} + r_1 + r_2 + r_3 + r_4 + r_5 +
{2 \over 3} \ r_6 + {1 \over 3} \ r_7 + {1 \over 3} \ r_8  \cr
\theta_2(\Lambda^+) &\equiv
{16 \over 3} + r_2 + r_3 + r_4 + r_5 + {2 \over 3} \ r_6 +
{1 \over 3} \ r_7 + {1 \over 3} \ r_8  \cr
\theta_3(\Lambda^+) &\equiv
{13 \over 3} + r_3 + r_4 + r_5 + {2 \over 3} \ r_6 +
{1 \over 3} \ r_7 + { 1 \over 3} \ r_8       \cr
\theta_4(\Lambda^+) &\equiv
{10 \over 3} + r_4 + r_5 + {2 \over 3} \ r_6 + {1 \over 3} \ r_7 +
{1 \over 3} \ r_8      \cr
\theta_5(\Lambda^+) &\equiv {7 \over 3} + r_5 + {2 \over 3} \ r_6 +
{1 \over 3} \ r_7 + {1 \over 3} \ r_8 \cr
\theta_6(\Lambda^+) &\equiv {4 \over 3} + {2 \over 3} \ r_6 +
{1 \over 3} \ r_7 + {1 \over 3} \ r_8  \cr
\theta_7(\Lambda^+) &\equiv {1 \over 3} - {1 \over 3} \ r_6 +
{1 \over 3} \ r_7 + {1 \over 3} \ r_8  \cr
\theta_8(\Lambda^+) &\equiv
- {2 \over 3} - {1 \over 3} \ r_6 - {2 \over 3} \ r_7 + {1 \over 3} \ r_8 \cr
\theta_9(\Lambda^+) &\equiv
- {68 \over 3} - r_1 - 2 \ r_2 - 3 \ r_3 - 4 \ r_4 - 5 \ r_5 -
{10 \over 3} \  r_6 - {5 \over 3} \ r_7 - {8 \over 3} \ r_8  } \ \ . \eqno(IV.6) $$
The parameters $r_i$ here are the ones introduced in (II.6) for a dominant
weight $\Lambda^+$. We notice here that $A_8$ dualities are valid also for
the basis functions ${\it \Theta}(M, \Lambda^+)$ because
${\it \Theta}(1, \Lambda^+) \equiv 0 $. This highly facilitates the work by
allowing us to decompose all invariant polinomials $ P_\alpha(\Lambda^+) $
in terms of ${\it \Theta}(M,\Lambda^+)$'s but only for M=2,3,..9.

$E_8$ cohomology manifests itself by the fact that for the degree 8
we have 2 polinomials ${\cal K}_1$ and ${\cal K}_2$ as {\bf $E_8$
Basis functions} in spite of the fact that there could be a priori
7 polinomials related with $A_8$ cohomology. For any 8th order Casimir
operator of $E_8$, the eigenvalues can always be expressed as linear
superpositions of these $E_8$ basis functions.

We will summarize our results for degrees 12 and 14 in appendix(4). As is
mentioned above, $A_8$ a priori give 19 and 29 polinomials for degrees 12
and 14 respectively. It is however seen that the cohomology of $E_8$ dictates
only 8 and 19 invariant polinomials for these degrees.

\vskip 3mm
\noindent {\bf{V.WEIGHT MULTIPLICITY FORMULAS FOR $E_8$ }}
\vskip 3mm
Careful reader could now raise the question that is there a way for a direct
comparison of our results in presenting the $E_8$ basis functions
$$ \eqalign{
&{\cal K}_\alpha(8,\Lambda^+) \ for  \ \alpha = 1,2 \cr
&{\cal K}_\alpha(12,\Lambda^+) \ for \ \alpha = 1,2 .. 8 \cr
&{\cal K}_\alpha(14,\Lambda^+) \ for \ \alpha = 1,2 .. 19 \ \ . }  $$
A simple and {\bf might be possible} way for such an investigation is due
to weight multiplicity formulas which can be obtained from these polinomials.
The method has been presented in another work {\bf [7]} for $A_N$ Lie algebras
and it can be applied here just as in the same manner. This shows the
correctness in our conclusion that any Casimir operator for $E_8$ can be
expressed as linear superpositions of $E_8$ basis functions which are given
in this work. An explicit comparison has been given in our previous works
but only for 4th and 5th order Casimir operators of $A_N$ Lie algebras and
beyond these this could be useless in practice.

On the other hand, one could give our present results beyond order 14,
namely for 18,20,24 and 30. This however could be a task for further
investigations. As the final remark, one can see that the method presented
in this paper are to be extended in the same manner to cases $E_7$ and $G_2$
in terms of their sub-groups $A_7$ and $A_2$.

\vskip 15mm
\vskip 15mm

\hfill\eject

\noindent {\bf{REFERENCES}}
\vskip 3mm
[1] R. Hermann : Chapter 10, Lie Groups for Physicists, (1966) Benjamin

[2] A. Borel and C. Chevalley : Mem.Am.Math.Soc. 14 (1955) 1

Chih-Ta Yen: Sur Les Polynomes de Poincare des Groupes de Lie Exceptionnels,
Comptes Rendue Acad.Sci. Paris (1949) 628-630

C. Chevalley : The Betti Numbers of the Exceptional Simple Lie Groups,
Proceedings of the International Congress of Mathematicians, 2 (1952) 21-24

A. Borel : Ann.Math. 57 (1953) 115-207

A.J. Coleman : Can.J.Math 10 (1958) 349-356

[3] H.R.Karadayi and M.Gungormez : Explicit Construction of Casimir
Operators and Eigenvalues:I , hep-th/9609060

H.R.Karadayi and M.Gungormez : Explicit Construction of Casimir
Operators and Eigenvalues:II , physics/9611002

[4] V.G.Kac, R.V.Moody and M.Wakimoto ; On $E_{10}$, preprint

[5] R.H Capps : The Weyl Orbits of $G_2$ , $F_4$ , $E_6$ and $E_7$,
Jour.Physics

R.H Capps : Geometrical Classification Scheme for Weights of Lie Algebras,
Jour.Math.Phys. 29 (1988) 1732,1735

[6] Humphreys J.E: Introduction to Lie Algebras and Representation
Theory , Springer-Verlag (1972) N.Y.

[7] H.R.Karadayi ; Non-Recursive Multiplicity Formulas for $A_N$ Lie algebras,
physics/9611008

\vskip 15mm
\vskip 15mm
\vskip 15mm
\vskip 15mm
\vskip 15mm
\vskip 15mm

\hfill\eject

\noindent {\bf{APPENDIX.1}}
\vskip 3mm
The Weyl orbits of $E_8$ fundamental dominant weights $\lambda_i$ \ (i=1,2, .. 8)
are the unions of those of the following $A_8$ dominant weights:
$$ \eqalign{
\Sigma(\lambda_2) \equiv  ( &\sigma_2 \ , \
2 \ \sigma_1 + \sigma_7 - \sigma_8 \ , \
\sigma_1 + \sigma_3 - \sigma_8   \cr
&\sigma_1 + \sigma_6 - \sigma_8 \ , \
\sigma_3 + \sigma_6 - 2 \sigma_8 \ , \
\sigma_5 + \sigma_7 - 2 \sigma_8 \ , \
\sigma_2 + \sigma_4 - 2 \ \sigma_8 )    \cr
&\ \ \ \ \ \ \ \ \ \ \ \ \ \ \ \ \ \ \
\ \ \ \ \ \ \ \ \ \ \ \ \ \ \ \ \ \ \
\ \ \ \ \ \ \ \ \ \ \ \ \ \ \ \ \ \ \cr
\Sigma(\lambda_3) \equiv  ( &\sigma_3 \ , \
\sigma_2 + \sigma_3 - \sigma_8 \ , \
\sigma_2 + \sigma_6 - \sigma_8 \cr
&3 \ \sigma_1 + \sigma_6 - 2 \ \sigma_8 \ , \
\sigma_4 + 2 \sigma_7 - 2 \sigma_8 \ , \
\sigma_1 + \sigma_3 + \sigma_6 - 2 \sigma_8 \cr
&2 \ \sigma_1 + \sigma_3 + \sigma_7 - 2 \ \sigma_8 \ , \
\sigma_1 + \sigma_5 + \sigma_7 - 2 \sigma_8 \ , \
2 \sigma_1 + \sigma_6 + \sigma_7 - 2 \sigma_8 \cr
&\sigma_1 + \sigma_2 + \sigma_4 - 2 \sigma_8 \ , \
\sigma_1 + 2 \sigma_4 - 3 \sigma_8 \ , \
2 \sigma_2 + \sigma_5 - 3 \sigma_8 \cr
&2 \sigma_5 - 3 \sigma_8 \ , \
\sigma_2 + \sigma_4 + \sigma_6 - 3 \sigma_8 \ , \
\sigma_3 + \sigma_5 + \sigma_7 - 3 \sigma_8 ) \cr
&\ \ \ \ \ \ \ \ \ \ \ \ \ \ \ \ \ \ \
\ \ \ \ \ \ \ \ \ \ \ \ \ \ \ \ \ \ \
\ \ \ \ \ \ \ \ \ \ \ \ \ \ \ \ \ \ \cr
\Sigma(\lambda_4) \equiv (&\sigma_4 \ , \
2 \ \sigma_3 - \sigma_8 \ , \
\sigma_3 + \sigma_6 - \sigma_8 \cr
&\sigma_1 + \sigma_4 + 2 \ \sigma_7 - 2 \ \sigma_8 \ , \
\sigma_2 + \sigma_3 + \sigma_6 - 2 \ \sigma_8 \ , \
\sigma_2 + \sigma_5 + \sigma_7 - 2 \ \sigma_8 \cr
&\sigma_3 + 3 \ \sigma_7 - 2 \ \sigma_8 \ , \
2 \ \sigma_2 + \sigma_4 - 2 \ \sigma_8 \ , \
4 \ \sigma_1 + \sigma_5 - 3 \ \sigma_8 \cr
&2 \ \sigma_1 + \sigma_3 + \sigma_6 + \sigma_7 - 3 \ \sigma_8 \ , \
2 \ \sigma_1 + 2 \ \sigma_4 - 3 \ \sigma_8 \ , \
\sigma_1 + 2 \ \sigma_2 + \sigma_5 - 3 \ \sigma_8 \cr
&\sigma_1 + 2 \sigma_5 - 3 \ \sigma_8 \ , \
3 \ \sigma_2 + \sigma_6 - 3 \ \sigma_8 \ , \
3 \ \sigma_1 + \sigma_3 + \sigma_6 - 3 \ \sigma_8 \cr
&\sigma_3 + \sigma_4 + 2 \ \sigma_7 - 3 \ \sigma_8 \ , \
2 \ \sigma_1 + \sigma_5 + 2 \ \sigma_7 - 3 \ \sigma_8 \ , \
\sigma_1 + \sigma_2 + \sigma_4 + \sigma_6 - 3 \ \sigma_8 \cr
&3 \ \sigma_1 + 2 \sigma_6 - 3 \ \sigma_8 \ , \
2 \sigma_1 + \sigma_2 + \sigma_4 + \sigma_7 - 3 \sigma_8 \ , \
\sigma_1 + \sigma_3 + \sigma_5 + \sigma_7 - 3 \sigma_8 \cr
&\sigma_2 + \sigma_4 + \sigma_5 + \sigma_7 - 4 \sigma_8 \ , \
3 \sigma_4 - 4 \sigma_8 \ , \
\sigma_3 + 2 \sigma_5 - 4 \sigma_8 \cr
&\sigma_1 + 2 \sigma_4 + \sigma_6 - 4 \sigma_8 \ , \
2 \sigma_2 + \sigma_5 + \sigma_6 - 4 \sigma_8 \ , \
3 \sigma_5 - 5 \sigma_8  ) \cr
&\ \ \ \ \ \ \ \ \ \ \ \ \ \ \ \ \ \ \
\ \ \ \ \ \ \ \ \ \ \ \ \ \ \ \ \ \ \
\ \ \ \ \ \ \ \ \ \ \ \ \ \ \ \ \ \ \cr
\Sigma(\lambda_5) \equiv ( &\sigma_5 \ , \
\sigma_4 + \sigma_6 - \sigma_8 \ , \
\sigma_2 + \sigma_4 + 2 \ \sigma_7 - 2 \ \sigma_8 \cr
&2 \ \ \sigma_3 + \sigma_6 - 2 \ \sigma_8 \ , \
\sigma_2 + 4 \ \sigma_7 - 2 \ \sigma_8 \ , \
\sigma_3 + \sigma_5 + \sigma_7 - 2 \ \sigma_8 \cr
&\sigma_1 + \sigma_3 + 3 \ \sigma_7 - 2 \ \sigma_8 \ , \
5 \ \sigma_1 + \sigma_4 - 3 \ \sigma_8 \ , \
2 \ \sigma_3 + 3 \ \sigma_7 - 3 \ \sigma_8 \cr
&2 \ \sigma_1 + \sigma_4 + 3 \ \sigma_7 - 3 \ \sigma_8 \ , \
3 \ \sigma_2 + \sigma_5 - 3 \ \sigma_8 \ , \
\sigma_2 + 2 \ \sigma_5 - 3 \ \sigma_8 \cr
&\sigma_1 + 3 \ \sigma_2 + \sigma_6 - 3 \ \sigma_8 \ , \
\sigma_1 + \sigma_3 + \sigma_4 + 2 \ \sigma_7 - 3 \ \sigma_8 \ , \
2 \ \sigma_2 + \sigma_4 + \sigma_6 - 3 \ \sigma_8 \cr
&4 \ \sigma_2 + \sigma_7 - 3 \ \sigma_8 \ , \
\sigma_2 + \sigma_3 + \sigma_5 + \sigma_7 - 3 \ \sigma_8 \ , \
3 \ \sigma_2 + 2 \ \sigma_6 - 4 \ \sigma_8 \cr
&3 \ \sigma_1 + \sigma_3 + 2 \ \sigma_6 - 4 \ \sigma_8 \ , \
3 \ \sigma_1 + 2 \ \sigma_4 + \sigma_7 - 4 \ \sigma_8 \ , \
2 \ \sigma_1 + 2 \ \sigma_2 + \sigma_5 + \sigma_7 - 4 \ \sigma_8 \cr
&\sigma_1 + \sigma_2 + \sigma_4 + \sigma_5 + \sigma_7 - 4 \ \sigma_8 \ , \
4 \ \sigma_1 + \sigma_3 + \sigma_5 - 4 \ \sigma_8 \ , \
2 \ \sigma_1 + \sigma_2 + \sigma_4 + \sigma_6 + \sigma_7 - 4 \ \sigma_8 \cr
&\sigma_1 + \sigma_3 + 2 \ \sigma_5 - 4 \ \sigma_8 \ , \
3 \ \sigma_1 + \sigma_2 + \sigma_4 + \sigma_6 - 4 \ \sigma_8 \ , \
\sigma_2 + 2 \ \sigma_4 + 2 \ \sigma_7 - 4 \ \sigma_8 \cr
&2 \ \sigma_1 + \sigma_3 + \sigma_5 + 2 \ \sigma_7 - 4 \sigma_8 \ , \
2 \ \sigma_1 + 2 \ \sigma_4 + \sigma_6 - 4 \ \sigma_8 \ , \
\sigma_1 + 2 \sigma_2 + \sigma_5 + \sigma_6 - 4 \ \sigma_8 \cr
&3 \ \sigma_4 + \sigma_6 - 5 \ \sigma_8 \ , \
\sigma_1 + 2 \ \sigma_4 + \sigma_5 + \sigma_7 - 5 \ \sigma_8 \ , \
2 \ \sigma_2 + 2 \ \sigma_5 + \sigma_7 - 5 \ \sigma_8 \cr
&\sigma_2 + \sigma_4 + 2 \ \sigma_5 - 5 \ \sigma_8 \ , \
\sigma_3 + 3 \ \sigma_5 - 6 \ \sigma_8  )  } $$

\vskip 15mm
\vskip 15mm

\hfill\eject

$$ \eqalign{
\Sigma(\lambda_6) \equiv ( &\sigma_6 \ , \
\sigma_4 + \sigma_7 - \sigma_8 \ , \
3 \ \sigma_7 - \sigma_8 \cr
&\sigma_2 + 2 \sigma_7 - \sigma_8 \ , \
\sigma_1 + \sigma_3 + 2 \sigma_7 - 2 \ \sigma_8 \ , \
2 \ \sigma_2 + \sigma_6 - 2 \sigma_8 \cr
&\sigma_1 + 2 \sigma_2 + \sigma_7 - 2 \ \sigma_8 \ , \
\sigma_2 + \sigma_4 + \sigma_7 - 2 \ \sigma_8 \ , \
3 \ \sigma_2 - 2 \ \sigma_8 \cr
&3 \ \sigma_1 + \sigma_3 - 2 \ \sigma_8 \ , \
\sigma_3 + \sigma_5 - 2 \ \sigma_8 \ , \
2 \ \sigma_1 + \sigma_2 + \sigma_5 - 3 \ \sigma_8 \cr
&\sigma_1 + \sigma_4 + \sigma_5 - 3 \ \sigma_8 \ , \
2 \ \sigma_1 + \sigma_4 + \sigma_6 - 3 \ \sigma_8 \ , \
2 \ \sigma_4 + \sigma_7 - 3 \ \sigma_8 \cr
&\sigma_1 + \sigma_2 + \sigma_5 + \sigma_7 - 3 \ \sigma_8 \ , \
\sigma_2 + 2 \sigma_5 - 4 \ \sigma_8 )  \cr
&\ \ \ \ \ \ \ \ \ \ \ \ \ \ \ \ \ \ \
\ \ \ \ \ \ \ \ \ \ \ \ \ \ \ \ \ \ \
\ \ \ \ \ \ \ \ \ \ \ \ \ \ \ \ \ \ \cr
\Sigma(\lambda_7) \equiv ( &\sigma_7 \ , \
\sigma_2 + \sigma_7 - \sigma_8 \ , \
\sigma_1 + \sigma_2 - \sigma_8 \ , \
\sigma_4 - \sigma_8 \ , \
\sigma_1 + \sigma_5 - 2 \ \sigma_8 ) \cr
&\ \ \ \ \ \ \ \ \ \ \ \ \ \ \ \ \ \ \
\ \ \ \ \ \ \ \ \ \ \ \ \ \ \ \ \ \ \
\ \ \ \ \ \ \ \ \ \ \ \ \ \ \ \ \ \ \cr
\Sigma(\lambda_8) \equiv ( &\sigma_8 \ , \
\sigma_3 + \sigma_7 - \sigma_8 \ , \
3 \ \sigma_1 - \sigma_8 \cr
&2 \ \sigma_2 - \sigma_8 \ , \
\sigma_5 - \sigma_8 \ , \
\sigma_1 + 2 \ \sigma_7 - \sigma_8 \cr
&\sigma_1 + \sigma_2 + \sigma_6 - 2 \ \sigma_8 \ , \
\sigma_1 + \sigma_4 + \sigma_7 - 2 \ \sigma_8 \ , \
2 \ \sigma_1 + \sigma_4 - 2 \ \sigma_8 \cr
&\sigma_2 + \sigma_5 - 2 \ \sigma_8 \ , \
\sigma_4 + \sigma_5 - 3 \ \sigma_8 ) }  $$

As an example of (II.10), let us calculate $\Sigma(\lambda_1+\lambda_7)$
which contains 15 elements in the form $\Sigma(\lambda_1)+\Sigma(\lambda_7)$.
Due to equality of square lengths with that of $\lambda_1+\lambda_7$ only the
following ones are valid:
$$ \eqalign{
\Sigma(\lambda_1+\lambda_7) \equiv ( &\sigma_1 + \sigma_7 \ , \
\sigma_1 + \sigma_2 + \sigma_7 - \sigma_8 \ , \
2 \ \sigma_1 + \sigma_2 - \sigma_8 \cr
&\sigma_1 + \sigma_4 - \sigma_8 \ , \
\sigma_6 + \sigma_7 - \sigma_8 \ , \
2 \ \sigma_1 + \sigma_5 - 2 \ \sigma_8 \cr
&\sigma_2 + \sigma_3 + \sigma_7 - 2 \ \sigma_8 \ , \
\sigma_1 + \sigma_2 + \sigma_3 - 2 \ \sigma_8 \ , \
\sigma_3 + \sigma_4 - 2 \ \sigma_8 \cr
&\sigma_2 + \sigma_6 + \sigma_7 - 2 \ \sigma_8 \ , \
\sigma_4 + \sigma_6 - 2 \ \sigma_8 \ , \
\sigma_1 + \sigma_3 + \sigma_5 - 3 \ \sigma_8 \ , \
\sigma_1 + \sigma_5 + \sigma_6 - 3 \ \sigma_8 ) \ \ . } $$

\vskip 3mm
\noindent {\bf{APPENDIX.2}}
\vskip 3mm
Let us first borrow, for a dominant weight $\sigma^+$ of $A_8$, the
following quantities from ref(3):
$$ \eqalign{
\Omega_8(\sigma^+) &\equiv \ 40320 \ Q(8)  \ \mu(8) \ + \cr
&20160 \ \bigl( \  \cr
& \ 35 \ Q(4,4)  \ \mu(4,4) +
14 \ Q(5,3)  \ \mu(5,3) +
7  \ Q(6,2)  \ \mu(6,2) +
2  \ Q(7,1)  \ \mu(7,1) \ \bigr) \ + \cr
&40320 \ \bigl( \                    \cr
& \ 20 \ Q(3,3,2)  \ \mu(3,3,2) +
15 \ Q(4,2,2)  \ \mu(4,2,2) \ + \cr
& \ 5  \ Q(4,3,1)  \ \mu(4,3,1) +
3  \ Q(5,2,1)  \ \mu(5,2,1) +
2  \ Q(6,1,1)  \ \mu(6,1,1) \ \bigr) \ + \cr
&13440 \ \bigl( \                        \cr
& \ 540 \ Q(2,2,2,2)  \ \mu(2,2,2,2) +
30  \ Q(3,2,2,1)  \ \mu(3,2,2,1) \  + \cr
& \ 40  \ Q(3,3,1,1)  \ \mu(3,3,1,1) +
15  \ Q(4,2,1,1)  \ \mu(4,2,1,1) +
18  \ Q(5,1,1,1) \ \mu(5,1,1,1) \ \bigr) \ + \cr
&483840 \ \bigl( \                          \cr
& \ 3 \ Q(2,2,2,1,1) \ \mu(2,2,2,1,1) + Q(3,2,1,1,1) \ \mu(3,2,1,1,1) +
2 \ Q(4,1,1,1,1)  \ \mu(4,1,1,1,1) \ \bigr) \ + \cr
&967680 \ \bigl( \                            \cr
& \ 3 \ Q(2,2,1,1,1,1)  \ \mu(2,2,1,1,1,1) +
5 \ Q(3,1,1,1,1,1)  \ \mu(3,1,1,1,1,1) \ \bigr) \ + \cr
&29030400 \ Q(2,1,1,1,1,1,1)  \ \mu(2,1,1,1,1,1,1) \ + \cr
&1625702400 \ Q(1,1,1,1,1,1,1,1)  \ \mu(1,1,1,1,1,1,1,1) } $$

$$ \eqalign{ \Omega_{12}(\Lambda^+) &\equiv 40320  \ Q(12)  \ \mu(12) \ + \cr
&5040 \ \bigl( \                         \cr
&1848  \ Q(6,6)  \ \mu(6,6) +
792   \ Q(7,5)  \ \mu(7,5) +
495   \ Q(8,4)  \ \mu(8,4) \ +  \cr
&220   \ Q(9,3)  \ \mu(9,3) +
66    \ Q(10,2)  \ \mu(10,2) +
12    \ Q(11,1)  \ \mu(11,1) \ \bigr) \ + \cr
&95040 \ \bigl( \                       \cr
&1575  \ Q(4,4,4)  \ \mu(4,4,4) +
210   \ Q(5,4,3)  \ \mu(5,4,3) +
252   \ Q(5,5,2)  \ \mu(5,5,2) \ + \cr
&280   \ Q(6,3,3)  \ \mu(6,3,3) +
105   \ Q(6,4,2)  \ \mu(6,4,2) +
42    \ Q(6,5,1)  \ \mu(6,5,1) \ + \cr
&60    \ Q(7,3,2)  \ \mu(7,3,2) +
30    \ Q(7,4,1)  \ \mu(7,4,1) +
45    \ Q(8,2,2)  \ \mu(8,2,2) \ + \cr
&15    \ Q(8,3,1)  \ \mu(8,3,1) +
5     \ Q(9,2,1)  \ \mu(9,2,1) +
2     \ Q(10,1,1)  \ \mu(10,1,1) \ \bigr) \ + \cr
&95040 \ \bigl( \                         \cr
&11200  \ Q(3,3,3,3)  \ \mu(3,3,3,3) \ + \cr
&700    \ Q(4,3,3,2)  \ \mu(4,3,3,2) +
1050   \ Q(4,4,2,2)  \ \mu(4,4,2,2) \ + \cr
&350    \ Q(4,4,3,1)  \ \mu(4,4,3,1) +
420    \ Q(5,3,2,2)  \ \mu(5,3,2,2) \ + \cr
&280    \ Q(5,3,3,1)  \ \mu(5,3,3,1) +
105    \ Q(5,4,2,1)  \ \mu(5,4,2,1) \ + \cr
&168    \ Q(5,5,1,1)  \ \mu(5,5,1,1) +
630    \ Q(6,2,2,2)  \ \mu(6,2,2,2) \ + \cr
&70     \ Q(6,3,2,1)  \ \mu(6,3,2,1) +
70     \ Q(6,4,1,1)  \ \mu(6,4,1,1) \ + \cr
&60     \ Q(7,2,2,1)  \ \mu(7,2,2,1) +
40     \ Q(7,3,1,1)  \ \mu(7,3,1,1) \ + \cr
&15     \ Q(8,2,1,1)  \ \mu(8,2,1,1) +
10     \ Q(9,1,1,1)  \ \mu(9,1,1,1) \ \bigr) \ + \cr
&380160 \ \bigl( \                          \cr
&1260  \ Q(3,3,2,2,2)  \ \mu(3,3,2,2,2) \ + \cr
&420   \ Q(3,3,3,2,1)  \ \mu(3,3,3,2,1) +
1890  \ Q(4,2,2,2,2)  \ \mu(4,2,2,2,2) \ + \cr
&105   \ Q(4,3,2,2,1)  \ \mu(4,3,2,2,1) +
140   \ Q(4,3,3,1,1)  \ \mu(4,3,3,1,1) \ + \cr
&105   \ Q(4,4,2,1,1)  \ \mu(4,4,2,1,1) +
189   \ Q(5,2,2,2,1)  \ \mu(5,2,2,2,1) \ + \cr
&42    \ Q(5,3,2,1,1)  \ \mu(5,3,2,1,1) +
63    \ Q(5,4,1,1,1)  \ \mu(5,4,1,1,1) \ + \cr
&42    \ Q(6,2,2,1,1)  \ \mu(6,2,2,1,1) +
42    \ Q(6,3,1,1,1)  \ \mu(6,3,1,1,1) \ + \cr
&18    \ Q(7,2,1,1,1)  \ \mu(7,2,1,1,1) +
18    \ Q(8,1,1,1,1)  \ \mu(8,1,1,1,1) \ \bigr) \ + \cr
&570240 \ \bigl( \                           \cr
&56700  \ Q(2,2,2,2,2,2)  \ \mu(2,2,2,2,2,2) \ + \cr
&1260   \ Q(3,2,2,2,2,1)  \ \mu(3,2,2,2,2,1) +
280    \ Q(3,3,2,2,1,1)  \ \mu(3,3,2,2,1,1) \ + \cr
&840    \ Q(3,3,3,1,1,1)  \ \mu(3,3,3,1,1,1) +
315    \ Q(4,2,2,2,1,1)  \ \mu(4,2,2,2,1,1) \ + \cr
&105    \ Q(4,3,2,1,1,1)  \ \mu(4,3,2,1,1,1) +
420    \ Q(4,4,1,1,1,1)  \ \mu(4,4,1,1,1,1) \ + \cr
&126    \ Q(5,2,2,1,1,1)  \ \mu(5,2,2,1,1,1) +
168    \ Q(5,3,1,1,1,1)  \ \mu(5,3,1,1,1,1) \ +  \cr
&84     \ Q(6,2,1,1,1,1)  \ \mu(6,2,1,1,1,1) +
120    \ Q(7,1,1,1,1,1)  \ \mu(7,1,1,1,1,1) \ \bigr) \ + \cr
&79833600 \ \bigl( \                                  \cr
&90  \ Q(2,2,2,2,2,1,1)  \ \mu(2,2,2,2,2,1,1) \ + \cr
&9   \ Q(3,2,2,2,1,1,1)  \ \mu(3,2,2,2,1,1,1) +
8   \ Q(3,3,2,1,1,1,1)  \ \mu(3,3,2,1,1,1,1) \ + \cr
&6   \ Q(4,2,2,1,1,1,1)  \ \mu(4,2,2,1,1,1,1) +
10  \ Q(4,3,1,1,1,1,1)  \ \mu(4,3,1,1,1,1,1) \ + \cr
&6   \ Q(5,2,1,1,1,1,1)  \ \mu(5,2,1,1,1,1,1) +
12  \ Q(6,1,1,1,1,1,1)  \ \mu(6,1,1,1,1,1,1) \ \bigr) \ + \cr
&479001600 \ \bigl( \                                  \cr
&36 \ Q(2,2,2,2,1,1,1,1)  \ \mu(2,2,2,2,1,1,1,1) \ + \cr
&10  \ Q(3,2,2,1,1,1,1,1)  \ \mu(3,2,2,1,1,1,1,1) +
40  \ Q(3,3,1,1,1,1,1,1)  \ \mu(3,3,1,1,1,1,1,1) \ + \cr
&15  \ Q(4,2,1,1,1,1,1,1)  \ \mu(4,2,1,1,1,1,1,1) +
42  \ Q(5,1,1,1,1,1,1,1)  \ \mu(5,1,1,1,1,1,1,1) \ \bigr) } $$

$$ \eqalign{ \Omega_{14}(\Lambda^+) &\equiv 40320 \ Q(14) \ \mu(14) \ +  \cr
&5040 \ ( \               \cr
&6864 \ Q(7,7) \ \mu(7,7) +
3003  \ Q(8,6)  \ \mu(8,6) +
2002  \ Q(9,5)  \ \mu(9,5) \ + \cr
&1001  \ Q(10,4)  \ \mu(10,4) +
364   \ Q(11,3)  \ \mu(11,3) +
91    \ Q(12,2)  \ \mu(12,2) +
14    \ Q(13,1)  \ \mu(13,1) \ ) \ + \cr
&65520 \ ( \                       \cr
&5544 \ Q(5,5,4) \ \mu(5,5,4) +
4620 \ Q(6,4,4) \ \mu(6,4,4) \ + \cr
&1848 \ Q(6,5,3) \ \mu(6,5,3) +
1848 \ Q(6,6,2)  \ \mu(6,6,2) +
1320 \ Q(7,4,3)  \ \mu(7,4,3) \ + \cr
&792 \ Q(7,5,2)  \ \mu(7,5,2) +
264 \ Q(7,6,1)  \ \mu(7,6,1) +
1320 \ Q(8,3,3)  \ \mu(8,3,3) \ + \cr
&495   \ Q(8,4,2)  \ \mu(8,4,2) +
198   \ Q(8,5,1)  \ \mu(8,5,1) +
220   \ Q(9,3,2)  \ \mu(9,3,2) \ +  \cr
&110   \ Q(9,4,1)  \ \mu(9,4,1) +
132   \ Q(10,2,2)  \ \mu(10,2,2) +
44    \ Q(10,3,1)  \ \mu(10,3,1) \ + \cr
&12    \ Q(11,2,1)  \ \mu(11,2,1) +
4     \ Q(12,1,1)  \ \mu(12,1,1) \ ) \ + \cr
&43680 \ ( \                      \cr
&46200  \ Q(4,4,3,3)  \ \mu(4,4,3,3) +
51975  \ Q(4,4,4,2)  \ \mu(4,4,4,2) \ + \cr
&55440  \ Q(5,3,3,3)  \ \mu(5,3,3,3) +
6930   \ Q(5,4,3,2)  \ \mu(5,4,3,2) +
6930   \ Q(5,4,4,1)  \ \mu(5,4,4,1) \ + \cr
&16632  \ Q(5,5,2,2)  \ \mu(5,5,2,2) +
5544   \ Q(5,5,3,1)  \ \mu(5,5,3,1) +
9240   \ Q(6,3,3,2)  \ \mu(6,3,3,2) \ + \cr
&6930   \ Q(6,4,2,2)  \ \mu(6,4,2,2) +
2310   \ Q(6,4,3,1)  \ \mu(6,4,3,1) +
1386   \ Q(6,5,2,1)  \ \mu(6,5,2,1) \ + \cr
&1848   \ Q(6,6,1,1)  \ \mu(6,6,1,1) +
3960   \ Q(7,3,2,2)  \ \mu(7,3,2,2) +
2640   \ Q(7,3,3,1)  \ \mu(7,3,3,1) \ + \cr
&990    \ Q(7,4,2,1)  \ \mu(7,4,2,1) +
792    \ Q(7,5,1,1)  \ \mu(7,5,1,1) +
4455   \ Q(8,2,2,2)  \ \mu(8,2,2,2) \ + \cr
&495    \ Q(8,3,2,1)  \ \mu(8,3,2,1) +
495    \ Q(8,4,1,1)  \ \mu(8,4,1,1) +
330    \ Q(9,2,2,1)  \ \mu(9,2,2,1) \ + \cr
&220    \ Q(9,3,1,1)  \ \mu(9,3,1,1) +
66     \ Q(10,2,1,1)  \ \mu(10,2,1,1) +
36     \ Q(11,1,1,1)  \ \mu(11,1,1,1) \ ) \ + \cr
&1153152 \ ( \                         \cr
&16800  \ Q(3,3,3,3,2)  \ \mu(3,3,3,3,2) \ + \cr
&2100   \ Q(4,3,3,2,2)  \ \mu(4,3,3,2,2) +
2100   \ Q(4,3,3,3,1)  \ \mu(4,3,3,3,1)  \ + \cr
&4725   \ Q(4,4,2,2,2)  \ \mu(4,4,2,2,2) +
525    \ Q(4,4,3,2,1)  \ \mu(4,4,3,2,1)  \ + \cr
&1575   \ Q(4,4,4,1,1)  \ \mu(4,4,4,1,1) +
1890   \ Q(5,3,2,2,2)  \ \mu(5,3,2,2,2)  \ + \cr
&420    \ Q(5,3,3,2,1)  \ \mu(5,3,3,2,1) +
315    \ Q(5,4,2,2,1)  \ \mu(5,4,2,2,1)  \ + \cr
&210    \ Q(5,4,3,1,1)  \ \mu(5,4,3,1,1) +
252    \ Q(5,5,2,1,1)  \ \mu(5,5,2,1,1)  \ + \cr
&3780   \ Q(6,2,2,2,2)  \ \mu(6,2,2,2,2)  +
210    \ Q(6,3,2,2,1)  \ \mu(6,3,2,2,1)  \ + \cr
&280    \ Q(6,3,3,1,1)  \ \mu(6,3,3,1,1) +
105    \ Q(6,4,2,1,1)  \ \mu(6,4,2,1,1)  \ + \cr
&126    \ Q(6,5,1,1,1)  \ \mu(6,5,1,1,1) +
270    \ Q(7,2,2,2,1)  \ \mu(7,2,2,2,1)  \ + \cr
&60     \ Q(7,3,2,1,1)  \ \mu(7,3,2,1,1)  +
90     \ Q(7,4,1,1,1)  \ \mu(7,4,1,1,1)  \ + \cr
&45     \ Q(8,2,2,1,1)  \ \mu(8,2,2,1,1) +
45     \ Q(8,3,1,1,1)  \ \mu(8,3,1,1,1)  \ + \cr
&15     \ Q(9,2,1,1,1)  \ \mu(9,2,1,1,1) +
12     \ Q(10,1,1,1,1)  \ \mu(10,1,1,1,1) \ ) \ + \cr
&1729728 \ ( \                                  \cr
&25200  \ Q(3,3,2,2,2,2)  \ \mu(3,3,2,2,2,2) +
4200   \ Q(3,3,3,2,2,1)  \ \mu(3,3,3,2,2,1)    \ + \cr
&11200  \ Q(3,3,3,3,1,1)  \ \mu(3,3,3,3,1,1)  +
47250  \ Q(4,2,2,2,2,2)  \ \mu(4,2,2,2,2,2)    \ + \cr
&1575   \ Q(4,3,2,2,2,1)  \ \mu(4,3,2,2,2,1) +
700    \ Q(4,3,3,2,1,1)  \ \mu(4,3,3,2,1,1)    \ + \cr
&1050   \ Q(4,4,2,2,1,1)  \ \mu(4,4,2,2,1,1) +
1050   \ Q(4,4,3,1,1,1)  \ \mu(4,4,3,1,1,1)    \ + \cr
&3780   \ Q(5,2,2,2,2,1)  \ \mu(5,2,2,2,2,1) +
420    \ Q(5,3,2,2,1,1)  \ \mu(5,3,2,2,1,1)    \ + \cr
&840    \ Q(5,3,3,1,1,1)  \ \mu(5,3,3,1,1,1) +
315    \ Q(5,4,2,1,1,1)  \ \mu(5,4,2,1,1,1)    \ + \cr
&1008   \ Q(5,5,1,1,1,1)  \ \mu(5,5,1,1,1,1) +
630    \ Q(6,2,2,2,1,1)  \ \mu(6,2,2,2,1,1)    \ + \cr
&210    \ Q(6,3,2,1,1,1)  \ \mu(6,3,2,1,1,1) +
420    \ Q(6,4,1,1,1,1)  \ \mu(6,4,1,1,1,1)    \ + \cr
&180    \ Q(7,2,2,1,1,1)  \ \mu(7,2,2,1,1,1) +
240    \ Q(7,3,1,1,1,1)  \ \mu(7,3,1,1,1,1)    \ + \cr
&90     \ Q(8,2,1,1,1,1)  \ \mu(8,2,1,1,1,1) +
100    \ Q(9,1,1,1,1,1)  \ \mu(9,1,1,1,1,1) \ ) \ + } $$

$$ \eqalign{
&34594560 \ ( \                  \cr
&198450 \ Q(2,2,2,2,2,2,2)  \ \mu(2,2,2,2,2,2,2) \ + \cr
&3150   \ Q(3,2,2,2,2,2,1)  \ \mu(3,2,2,2,2,2,1) +
420     \ Q(3,3,2,2,2,1,1)  \ \mu(3,3,2,2,2,1,1)  \ + \cr
&420    \ Q(3,3,3,2,1,1,1)  \ \mu(3,3,3,2,1,1,1) +
630     \ Q(4,2,2,2,2,1,1)  \ \mu(4,2,2,2,2,1,1)  \ + \cr
&105    \ Q(4,3,2,2,1,1,1)  \ \mu(4,3,2,2,1,1,1) +
280     \ Q(4,3,3,1,1,1,1)  \ \mu(4,3,3,1,1,1,1)  \ + \cr
&210    \ Q(4,4,2,1,1,1,1)  \ \mu(4,4,2,1,1,1,1) +
189     \ Q(5,2,2,2,1,1,1)  \ \mu(5,2,2,2,1,1,1)  \ + \cr
&84     \ Q(5,3,2,1,1,1,1)  \ \mu(5,3,2,1,1,1,1) +
210     \ Q(5,4,1,1,1,1,1)  \ \mu(5,4,1,1,1,1,1)  \ + \cr
&84     \ Q(6,2,2,1,1,1,1)  \ \mu(6,2,2,1,1,1,1) +
140     \ Q(6,3,1,1,1,1,1)  \ \mu(6,3,1,1,1,1,1)  \ + \cr
&60     \ Q(7,2,1,1,1,1,1)  \ \mu(7,2,1,1,1,1,1) +
90      \ Q(8,1,1,1,1,1,1)  \ \mu(8,1,1,1,1,1,1) \ ) \ + \cr
&7264857600 \ ( \                          \cr
&270  \ Q(2,2,2,2,2,2,1,1)  \ \mu(2,2,2,2,2,2,1,1) \ + \cr
&18   \ Q(3,2,2,2,2,1,1,1)  \ \mu(3,2,2,2,2,1,1,1) +
8 \ Q(3,3,2,2,1,1,1,1)  \ \mu(3,3,2,2,1,1,1,1)   \ + \cr
&40 \ Q(3,3,3,1,1,1,1,1)  \ \mu(3,3,3,1,1,1,1,1) +
9 \ Q(4,2,2,2,1,1,1,1)  \ \mu(4,2,2,2,1,1,1,1)   \ + \cr
&5 \ Q(4,3,2,1,1,1,1,1)  \ \mu(4,3,2,1,1,1,1,1) +
30 \ Q(4,4,1,1,1,1,1,1)  \ \mu(4,4,1,1,1,1,1,1)  \ + \cr
&6 \ Q(5,2,2,1,1,1,1,1)  \ \mu(5,2,2,1,1,1,1,1) +
12 \ Q(5,3,1,1,1,1,1,1)  \ \mu(5,3,1,1,1,1,1,1)  \ + \cr
&6    \ Q(6,2,1,1,1,1,1,1)  \ \mu(6,2,1,1,1,1,1,1) +
12   \ Q(7,1,1,1,1,1,1,1)  \ \mu(7,1,1,1,1,1,1,1)) \ }  $$
In all these expressions, the so-called Q-generators are to be reduced
to the ones defined by (III.3) for which the parameters $q_i$ are determined
via (III.1) for a $\sigma^+$ which we prefer to suppress from Q-generators.
The reduction rules can be deduced from definitions given also in ref(3).
The left-hand side of (III.2) can thus be calculated from
$$ ch_M(\sigma^+) \equiv {1 \over 9!} \ dim\Pi(\sigma^+) \ \Omega_M(\sigma^+) $$
with which we obtain $E_8$ Weyl orbit characters. As is given in ref(3), the
number of elements which are contained within any $A_8$ Weyl orbit
$\Pi(\sigma^+)$ is $dim\Pi(\sigma^+)$ and it is known explicitly due to
permutational lemma for any $\sigma^+$.

\vskip 3mm
\noindent {\bf{APPENDIX.3}}
\vskip 3mm
In addition to the one given in (III.5), $A_8$ dualities give rise to the
following non-linear dependences for $\mu$-generators

$$ \eqalign{
\mu(11) &\equiv {1  \over 362880} \ ( \ - 3465  \ \mu(2)^4 \ \mu(3) \cr
&+ 12320   \ \mu(2)    \ \mu(3)^3  +
41580   \ \mu(2)^2  \ \mu(3)    \ \mu(4)             \cr
&- 41580   \ \mu(3)    \ \mu(4)^2            +
16632   \ \mu(2)^3  \ \mu(5)                         \cr
&- 44352   \ \mu(3)^2  \ \mu(5)              -
99792   \ \mu(2)    \ \mu(4)    \ \mu(5)             \cr
&- 110880  \ \mu(2)    \ \mu(3)    \ \mu(6)   +
133056  \ \mu(5)    \ \mu(6)                         \cr
&- 71280   \ \mu(2)^2  \ \mu(7)              +
142560  \ \mu(4)    \ \mu(7)                         \cr
&+ 166320  \ \mu(3)    \ \mu(8)              +
221760  \ \mu(2)    \ \mu(9) \ )                    } $$

$$ \eqalign{
\mu(12) &\equiv {1 \over 725760} \ ( \ 322560  \ \mu(3) \ \mu(9) \cr
&+ 136080  \ \mu(2)^2  \ \mu(8)              +
272160  \ \mu(4)    \ \mu(8)                    \cr
&+ 248832  \ \mu(5)    \ \mu(7)              -
60480   \ \mu(2)^3  \ \mu(6)                    \cr
&- 80640   \ \mu(3)^2  \ \mu(6)             +
120960  \ \mu(6)^2                              \cr
&- 72576   \ \mu(2)^2  \ \mu(3)    \ \mu(5)  -
145152  \ \mu(3)    \ \mu(4)    \ \mu(5)        \cr
&+ 17010   \ \mu(2)^4  \ \mu(4)             -
34020   \ \mu(2)^2  \ \mu(4)^2                  \cr
&- 22680   \ \mu(4)^3                       +
20160   \ \mu(2)^3  \ \mu(3)^2                  \cr
&+ 4480    \ \mu(3)^4                       -
945     \ \mu(2)^6  \ )                         \cr
& \ \ \ \ \ \ \ \ \                                  \cr
\mu(14) &\equiv {1 \over 8709120} \ ( \ - 2835 \ \mu(2)^7  \cr
&+ 17640    \ \mu(2)^4  \ \mu(3)^2                     +
125440   \ \mu(2)    \ \mu(3)^4                           \cr
&+ 39690    \ \mu(2)^5  \ \mu(4)                       +
635040   \ \mu(2)^2  \ \mu(3)^2  \ \mu(4)                 \cr
&+ 79380    \ \mu(2)^3  \ \mu(4)^2                      -
635040   \ \mu(3)^2  \ \mu(4)^2                           \cr
&- 476280   \ \mu(2)    \ \mu(4)^3                      -
301056   \ \mu(3)^3  \ \mu(5)                             \cr
&- 2032128  \ \mu(2)    \ \mu(3)    \ \mu(4)  \ \mu(5)    -
158760   \ \mu(2)^4  \ \mu(6)                             \cr
&- 1128960  \ \mu(2)    \ \mu(3)^2  \ \mu(6)             -
635040   \ \mu(2)^2  \ \mu(4)    \ \mu(6)                 \cr
&+ 635040   \ \mu(4)^2  \ \mu(6)                        -
725760   \ \mu(2)^2  \ \mu(3)    \ \mu(7)                 \cr
&+ 1451520  \ \mu(3)    \ \mu(4)    \ \mu(7)             +
1244160  \ \mu(7)^2                                       \cr
&+ 317520   \ \mu(2)^3  \ \mu(8)                        +
846720   \ \mu(3)^2  \ \mu(8)                             \cr
&+ 1905120  \ \mu(2)    \ \mu(4)    \ \mu(8)             +
2540160  \ \mu(6)    \ \mu(8)                             \cr
&+ 2257920  \ \mu(2)    \ \mu(3)    \ \mu(9)             +
2709504  \ \mu(5)    \ \mu(9) \ )  } $$

\vskip 3mm
\noindent {\bf{APPENDIX.4}}
\vskip 3mm
We now give our results for degrees 12 and 14. Explicit dependences on
$\Lambda^+$ will be suppressed here. It will be useful to introduce the
following auxiliary functions in terms of which the formal definitions
of $E_8$ basis functions will be highly simplified:
$$ \eqalign{
{\cal W}_1(8) &\equiv 68580  \ {\it \Theta}(8)
- 42672  \ {\it \Theta}(2)  \ {\it \Theta}(6) \ -  \cr
&42672  \ {\it \Theta}(3)  \ {\it \Theta}(5)
- 13335  \ {\it \Theta}(4)^2    \ +      \cr
&13335  \ {\it \Theta}(2)^2  \ {\it \Theta}(4)
+ 17780  \ {\it \Theta}(2)  \ {\it \Theta}(3)^2
- 939    \ {\it \Theta}(2)^4   \cr
& \ \ \ \ \ \ \ \ \ \ \       \cr
{\cal W}_2(8) &\equiv  76765890960  \ {\it \Theta}(8)
- 47741514624  \ {\it \Theta}(2)  \ {\it \Theta}(6)   \ -  \cr
&47569228416  \ {\it \Theta}(3)  \ {\it \Theta}(5)
- 14950629660  \ {\it \Theta}(4)^2   \ +    \cr
&14921466630  \ {\it \Theta}(2)^2  \ {\it \Theta}(4)
+ 19832476160  \ {\it \Theta}(2)  \ {\it \Theta}(3)^2
- 1050561847   \ {\it \Theta}(2)^4  }  $$
$$ \eqalign{
{\cal W}_1(12) &\equiv 302400 \ {\it \Theta}(3) \ {\it \Theta}(9) -
56700   \ {\it \Theta}(4)  \ {\it \Theta}(8)     \ -    \cr
&51840   \ {\it \Theta}(5)  \ {\it \Theta}(7)
- 158400  \ {\it \Theta}(2)  \ {\it \Theta}(3)  \ {\it \Theta}(7)
+ 30240   \ {\it \Theta}(6)^2       \ -     \cr
&168000  \ {\it \Theta}(3)^2  \ {\it \Theta}(6)
+ 33264   \ {\it \Theta}(2)  \ {\it \Theta}(5)^2
- 80640   \ {\it \Theta}(3)  \ {\it \Theta}(4)  \ {\it \Theta}(5)  \ +  \cr
&16275   \ {\it \Theta}(4)^3
+ 92400   \ {\it \Theta}(2)  \ {\it \Theta}(3)^2  \ {\it \Theta}(4)
+ 19600   \ {\it \Theta}(3)^4   }  $$
$$ \eqalign{
{\cal W}_2(12) &\equiv 42338419200   \ {\it \Theta}(3)  \ {\it \Theta}(9)
- 7938453600    \ {\it \Theta}(4)  \ {\it \Theta}(8)  \ - \cr
&250343238600  \ {\it \Theta}(2)^2  \ {\it \Theta}(8)
- 7258014720    \ {\it \Theta}(5)  \ {\it \Theta}(7) \ -  \cr
&22177267200   \ {\it \Theta}(2)  \ {\it \Theta}(3)  \ {\it \Theta}(7)
+ 4233841920    \ {\it \Theta}(6)^2     \ -         \cr
&23521344000   \ {\it \Theta}(3)^2  \ {\it \Theta}(6)
+ 156357159840  \ {\it \Theta}(2)^3  \ {\it \Theta}(6)   \ +          \cr
&4657226112    \ {\it \Theta}(2)  \ {\it \Theta}(5)^2
- 11290245120   \ {\it \Theta}(3)  \ {\it \Theta}(4)  \ {\it \Theta}(5)   \ +    \cr
&160591001760  \ {\it \Theta}(2)^2  \ {\it \Theta}(3)  \ {\it \Theta}(5)
+ 2278630200    \ {\it \Theta}(4)^3      \ +     \cr
&48089818350   \ {\it \Theta}(2)^2  \ {\it \Theta}(4)^2
+ 12936739200   \ {\it \Theta}(2)  \ {\it \Theta}(3)^2  \ {\it \Theta}(4)  \ -   \cr
&48806484300   \ {\it \Theta}(2)^4  \ {\it \Theta}(4)
+ 2744156800    \ {\it \Theta}(3)^4        \ -    \cr
&66618900600   \ {\it \Theta}(2)^3  \ {\it \Theta}(3)^2
+ 3440480295    \ {\it \Theta}(2)^6    } $$
$$ \eqalign{
{\cal W}_3(12) &\equiv 1976486400  \ {\it \Theta}(3)  \ {\it \Theta}(9)
- 370591200   \ {\it \Theta}(4)  \ {\it \Theta}(8)      \ +          \cr
&63622800    \ {\it \Theta}(2)^2  \ {\it \Theta}(8)
- 338826240   \ {\it \Theta}(5)  \ {\it \Theta}(7)      \ -           \cr
&1035302400  \ {\it \Theta}(2)  \ {\it \Theta}(3)  \ {\it \Theta}(7)
+ 197648640   \ {\it \Theta}(6)^2                 \ -         \cr
&1098048000  \ {\it \Theta}(3)^2  \ {\it \Theta}(6)
- 12136320    \ {\it \Theta}(2)^3  \ {\it \Theta}(6)     \ +         \cr
&217413504   \ {\it \Theta}(2)  \ {\it \Theta}(5)^2
- 527063040   \ {\it \Theta}(3)  \ {\it \Theta}(4)  \ {\it \Theta}(5)    \ +    \cr
&185512320   \ {\it \Theta}(2)^2  \ {\it \Theta}(3)  \ {\it \Theta}(5)
+ 106373400   \ {\it \Theta}(4)^3            \ -         \cr
&39822300    \ {\it \Theta}(2)^2  \ {\it \Theta}(4)^2
+ 603926400   \ {\it \Theta}(2)  \ {\it \Theta}(3)^2  \ {\it \Theta}(4)   \ +  \cr
&6366150     \ {\it \Theta}(2)^4  \ {\it \Theta}(4)
+ 128105600   \ {\it \Theta}(3)^4              \ -     \cr
&63571200    \ {\it \Theta}(2)^3  \ {\it \Theta}(3)^2
- 274935      \ {\it \Theta}(2)^6       }  $$
$$ \eqalign{
{\cal W}_4(12) &\equiv - 1501985020838400  \ {\it \Theta}(3)  \ {\it \Theta}(9)
+ 192772901311200   \ {\it \Theta}(4)  \ {\it \Theta}(8)     \ +      \cr
&2407922770302000  \ {\it \Theta}(2)^2  \ {\it \Theta}(8)
- 13295642434560    \ {\it \Theta}(5)  \ {\it \Theta}(7)     \ +   \cr
&760428342950400   \ {\it \Theta}(2)  \ {\it \Theta}(3)  \ {\it \Theta}(7)
- 156516673824000   \ {\it \Theta}(6)^2       \ +      \cr
&33565287369600    \ {\it \Theta}(2)  \ {\it \Theta}(4)  \ {\it \Theta}(6)
+ 883577458444800   \ {\it \Theta}(3)^2  \ {\it \Theta}(6)   \ -      \cr
&1515778400455200  \ {\it \Theta}(2)^3  \ {\it \Theta}(6)
- 53070803904384    \ {\it \Theta}(2)  \ {\it \Theta}(5)^2   \ +      \cr
&579544204861440   \ {\it \Theta}(3)  \ {\it \Theta}(4)  \ {\it \Theta}(5)
- 1696086939738240  \ {\it \Theta}(2)^2  \ {\it \Theta}(3)  \ {\it \Theta}(5)  \ -  \cr
&47654628701400    \ {\it \Theta}(4)^3
- 461057612469300   \ {\it \Theta}(2)^2  \ {\it \Theta}(4)^2  \ -   \cr
&463327486742400   \ {\it \Theta}(2)  \ {\it \Theta}(3)^2  \ {\it \Theta}(4)
+ 472701971331450   \ {\it \Theta}(2)^4  \ {\it \Theta}(4)    \ -     \cr
&111245008649600   \ {\it \Theta}(3)^4
+ 684206487048000   \ {\it \Theta}(2)^3  \ {\it \Theta}(3)^2
- 33351005297925    \ {\it \Theta}(2)^6      }   $$
$$ \eqalign{
{\cal W}_1(14) &\equiv 211680  \ {\it \Theta}(9) \ {\it \Theta}(5)
+ 26460   \ {\it \Theta}(8) \ {\it \Theta}(6)
- 43200   \ {\it \Theta}(7)^2 \ -   \cr
&29400   \ {\it \Theta}(8) \ {\it \Theta}(3)^2
+ 58800   \ {\it \Theta}(7) \ {\it \Theta}(4) \ {\it \Theta}(3)
- 122304  \ {\it \Theta}(6) \ {\it \Theta}(5) \ {\it \Theta}(3) \ -  \cr
&12495   \ {\it \Theta}(6) \ {\it \Theta}(4)^2
- 91728   \ {\it \Theta}(5)^2  \ {\it \Theta}(4)
+ 27440   \ {\it \Theta}(5) \ {\it \Theta}(3)^3
- 9800    \ {\it \Theta}(4)^2  \ {\it \Theta}(3)^2  }  $$
It is first seen that the expression (IV.4) can be cast in the form
$$ {\cal K}_2(8) \equiv {\cal W}_1(8) \ + \ 385526887200 \ \  . $$
Let us further define
$$ \eqalign{
{\cal Q}_{12} &\equiv ( \ {\it \Theta}(2) - 620) \ ( \ - 105 \ {\it \Theta}(2)^5
+ 341250             \ {\it \Theta}(2)^4
- 443786280          \ {\it \Theta}(2)^3   \ + \cr
&288672359200       \ {\it \Theta}(2)^2
- 93922348435072     \ {\it \Theta}(2)
+ 12228055880335360  \  )  } $$
and
$$ \eqalign{
{\cal Q}_{14} &\equiv ( \ {\it \Theta}(2) - 620) \ ( \ - 3 \ {\it \Theta}(2)^6
+ 11790               \ {\it \Theta}(2)^5
- 19314252            \ {\it \Theta}(2)^4
+ 16882085360         \ {\it \Theta}(2)^3 \ -  \cr
&8303952287424       \ {\it \Theta}(2)^2
+ 2179380420445440    \ {\it \Theta}(2)
- 238431403767424000 \ )  }  $$
with the remark that the square length of $E_8$ Weyl vector is 620. At last,
8 and 19 $E_8$ basis functions will be expressed as in the following:
$$ \eqalign{
{\cal K}_1(12) &\equiv {\cal W}_1(12) \ + \ \cr
& {_{105} \over _{1392517035128}} \ ( \   \cr
&_{2327783} \ {\cal W}_2(8) \ {\it \Theta}(2)^2 \ + \ \cr
&_{1641651348800} \ {\cal W}_1(8) \ {\it \Theta}(2)^2 \ + \ \cr
&_{1853819288565353101504512} \ {\it \Theta}(2)^2 \ - \ \cr
&_{5646385058438400} \ {\cal W}_1(8) \ {\it \Theta}(2) \ - \ \cr
&_{2457714965901036308812800000} \ {\it \Theta}(2) \ + \ \cr
&_{1878213525838949376} \ {\cal W}_1(8) \ +  \ \cr
&_{474462162108792} \ {\cal K}_{12}(0) \ +  \ \cr
&_{814849980464400425555898009600} \ ) \ \cr
& \ \ \ \ \ \ \ \ \ \ \ \ \          \cr
{\cal K}_2(12) &\equiv {\cal W}_1(12) \ + \ \cr
& {_{105} \over _{6580376}} \ ( \ \cr
&_{11} \ {\cal W}_2(8) \ {\it \Theta}(2)^2 \ - \cr
&_{13946970} \ {\cal W}_1(8) \ {\it \Theta}(2)^2 \ - \ \cr
&_{717386789108493504} \ {\it \Theta}(2)^2 \ +  \ \cr
&_{2185025300} \ {\cal W}_1(8) \ {\it \Theta}(2) \ +  \ \cr
&_{951080970408987600000} \ {\it \Theta}(2) \ - \ \cr
&_{726826815792} \ {\cal W}_1(8) \ - \ \cr
&_{315043889595739569446400} \  )  \cr
& \ \ \ \ \ \ \ \ \ \ \ \ \         }  $$
$$ \eqalign{
{\cal K}_3(12) &\equiv - {_{10742925608415} \over _{467309767}} \ {\cal Q}_{12}
+ {_{6983349} \over _{10867669}}  \  {\cal K}_1(12)
+ {_{3884320} \over _{10867669}}  \  {\cal K}_2(12) \cr
& \ \ \ \ \ \ \ \ \ \ \ \ \ \                   \cr
{\cal K}_4(12) &\equiv - {_{2898884687985} \over _{487195289}} \ {\cal Q}_{12}
+ {_{39572311} \over _{237932583}}  \ {\cal K}_1(12)
+ {_{198360272} \over _{237932583}} \ {\cal K}_2(12)   \cr
& \ \ \ \ \ \ \ \ \ \ \ \ \ \                   \cr
{\cal K}_5(12) &\equiv - {_{511567886115} \over _{1063875427}} \ {\cal Q}_{12}
+ {_{2327783} \over _{173189023}}   \ {\cal K}_1(12)
+ {_{170861240} \over _{173189023}} \ {\cal K}_2(12)   \cr
& \ \ \ \ \ \ \ \ \ \ \ \ \ \                   \cr
{\cal K}_6(12) &\equiv - {_{2557839430575} \over _{69599327}} \  {\cal Q}_{12}
+ {_{11638915} \over _{11330123}}  \  {\cal K}_1(12)
- {_{308792} \over _{11330123}}   \  {\cal K}_2(12)   \cr
& \ \ \ \ \ \ \ \ \ \ \ \ \ \                   \cr
{\cal K}_7(12) &\equiv - {_{17904876014025} \over _{1362158257}} \ {\cal Q}_{12}
+ {_{11638915} \over _{31678099}}         \ {\cal K}_1(12)
+ {_{20039184} \over _{31678099}}         \ {\cal K}_2(12) \cr
& \ \ \ \ \ \ \ \ \ \ \ \ \ \                   \cr
{\cal K}_8(12) &\equiv - {_{26924625585} \over _{9942761}} \ {\cal Q}_{12}
+ {_{2327783} \over _{30753191}}    \ {\cal K}_1(12)
+ {_{28425408} \over _{30753191}}   \ {\cal K}_2(12)  }    $$

\vskip 15mm

\hfill\eject

$$ \eqalign{
{\cal K}_1(14) &\equiv {\cal W}_1(14)  \  +            \cr
& {_{1} \over _{409396489250473267200}} \ ( \ -  \cr
&_{50198389200} \ {\cal W}_2(12) \  {\it \Theta}(2)      \           +  \cr
&_{1814183745255} \  {\cal W}_3(12) \  {\it \Theta}(2)   \            +  \cr
&_{719963} \  {\cal W}_4(12) \  {\it \Theta}(2)          \            +  \cr
&_{114941496614400} \  {\cal W}_2(12)                    \         -  \cr
&_{2663797055081400} \  {\cal W}_3(12)                   \         +  \cr
&_{282784300728374808115200} \  {\cal W}_1(8) \  {\it \Theta}(2)  \     -  \cr
&_{63215721507749516817408000} \  {\cal W}_1(8)                 \    -  \cr
&_{1088012169332650346400} \  {\it \Theta}(2)^7                 \    +  \cr
&_{4950455370463559076120000}  \ {\it \Theta}(2)^6              \    -  \cr
&_{9655764190981762706790537600} \  {\it \Theta}(2)^5           \    +  \cr
&_{10465560623426071788759214080000} \  {\it \Theta}(2)^4       \    -  \cr
&_{6807538538432150822570374332825600} \  {\it \Theta}(2)^3     \    +  \cr
&_{2657394331038056352474406872078336000} \  {\it \Theta}(2)^2  \    -  \cr
&_{576387923771111244952245791209505280000} \  {\it \Theta}(2)  \    +  \cr
&_{53583585232103605801009946012851814400000} \ )  }  $$
$$ \eqalign{
{\cal K}_2(14) &\equiv {\cal W}_1(14) + \     \cr
& {_{1} \over _{568635456614400}} \ ( \                      \      \cr
& {\cal W}_4(12)  \ {\it \Theta}(2)                        \      +  \cr
&_{19504800} \  {\cal W}_2(12)                             \        -  \cr
&_{10727283367475136000}  \ {\cal W}_1(8)                    \        +  \cr
&_{47986596584438400} \  {\cal W}_1(8) \  {\it \Theta}(2)    \           +  \cr
&_{16539581376126884275200}  \ {\it \Theta}(2)^3             \        -  \cr
&_{33154342667599799842560000} \  {\it \Theta}(2)^2          \        +  \cr
&_{22141768911037509306263040000} \  {\it \Theta}(2)         \        -  \cr
&_{4925899975161328995062246400000}    \  )        } $$
$$ \eqalign{
{\cal K}_3(14) &\equiv {\cal W}_1(14) + \    \cr
& {_{1} \over _{96046509070368460800}} \ ( \ \cr             \
&_{1639053360} \  {\cal W}_2(12) \  {\it \Theta}(2)        \           -  \cr
&_{7819677495} \  {\cal W}_3(12) \  {\it \Theta}(2)        \           +  \cr
&_{168907} \  {\cal W}_4(12)  \ {\it \Theta}(2)            \           +  \cr
&_{52014010923000}  \ {\cal W}_3(12)                       \        -  \cr
&_{2304112252773339559835020800000}  \ {\it \Theta}(2)       \        +  \cr
&_{1634892368046326720660996352000000}  \  )  }   $$

\vskip 15mm

\hfill\eject

$$ \eqalign{
{\cal K}_{4}(14) &\equiv
-{_{3518381271825} \over _{4968075880576}} {\cal Q}_{14}
+{_{341316459225} \over _{426944020987}}   {\cal K}_1(14)
-{_{522985910360} \over _{426944020987}}   {\cal K}_2(14)
+{_{608613472122} \over _{426944020987}}   {\cal K}_3(14)  \cr
& \ \ \ \ \ \ \ \ \       \cr
{\cal K}_{5}(14) &\equiv
+{_{1555746820875} \over _{12804209320064}}  {\cal Q}_{14}
-{_{150922243875} \over _{1100361738443}}    {\cal K}_1(14)
+{_{8647219074000} \over _{7702532169101}}   {\cal K}_2(14)
+{_{1668191078} \over _{114963166703}}       {\cal K}_3(14) \cr
& \ \ \ \ \ \ \ \ \       \cr
{\cal K}_{6}(14) &\equiv
-{_{6486267514725} \over _{15677800418432}}  {\cal Q}_{14}
+{_{629229662925} \over _{1347310973459}}    {\cal K}_1(14)
-{_{28104693257260} \over _{9431176814213}}  {\cal K}_2(14)
+{_{33131262430998} \over _{9431176814213}}  {\cal K}_3(14) \cr
& \ \ \ \ \ \ \ \ \       \cr
{\cal K}_{7}(14) &\equiv
-{_{674600188425} \over _{1456122839168}}    {\cal Q}_{14}
+{_{588983731225} \over _{1126220008419}}    {\cal K}_1(14)
-{_{33746326745200} \over _{7883540058933}}  {\cal K}_2(14)
+{_{186601893958} \over _{39221592333}}      {\cal K}_3(14)  \cr
& \ \ \ \ \ \ \ \ \       \cr
{\cal K}_{8}(14) &\equiv
-{_{6298570125} \over _{84286231424}}      {\cal Q}_{14}
+{_{11609403375} \over _{137623612247}}    {\cal K}_1(14)
-{_{2012756994830} \over _{8670287571561}} {\cal K}_2(14)
+{_{9951652153766} \over _{8670287571561}} {\cal K}_3(14)    \cr
& \ \ \ \ \ \ \ \ \       \cr
{\cal K}_{9}(14) &\equiv
-{_{15028681275} \over _{171940704256}}  {\cal Q}_{14}
+{_{1457925075} \over _{14776154272}}    {\cal K}_1(14)
-{_{5220816525} \over _{6464567494}}     {\cal K}_2(14)
+{_{2638218937} \over _{1543777312}}     {\cal K}_3(14)     \cr
& \ \ \ \ \ \ \ \ \       \cr
{\cal K}_{10}(14) &\equiv
-{_{9597761156475} \over _{31040670265216}}  {\cal Q}_{14}
+{_{931074150675} \over _{2667557600917}}    {\cal K}_1(14)
-{_{53346689979600} \over _{18672903206419}} {\cal K}_2(14)
+{_{977642897482} \over _{278700047857}}     {\cal K}_3(14)  \cr
& \ \ \ \ \ \ \ \ \       \cr
{\cal K}_{11}(14) &\equiv
-{_{378166150305} \over _{424558809856}} {\cal Q}_{14}
+{_{36685714665} \over _{36485522722}}   {\cal K}_1(14)
+{_{26114321216} \over _{383097988581}}  {\cal K}_2(14)
-{_{56432673235} \over _{766195977162}}  {\cal K}_3(14)     \cr
& \ \ \ \ \ \ \ \ \       \cr
{\cal K}_{12}(14) &\equiv
+{_{23430680865} \over _{381493437568}}     {\cal Q}_{14}
-{_{43186980555} \over _{622907253529}}     {\cal K}_1(14)
+{_{2474434996560} \over _{4360350774703}}  {\cal K}_2(14)
+{_{32660069284} \over _{65079862309}}      {\cal K}_3(14)  \cr
& \ \ \ \ \ \ \ \ \       \cr
{\cal K}_{13}(14) &\equiv
+{_{957382659} \over _{12109629824}}  {\cal Q}_{14}
-{_{92875227} \over _{1040671313}}    {\cal K}_1(14)
+{_{5321365584} \over _{7284699191}}  {\cal K}_2(14)
+{_{2613460196} \over _{7284699191}}  {\cal K}_3(14)      \cr
& \ \ \ \ \ \ \ \ \       \cr
{\cal K}_{14}(14) &\equiv
+{_{1029186358425} \over _{5031952172992}}   {\cal Q}_{14}
-{_{199681738050} \over _{864866779733}}     {\cal K}_1(14)
+{_{11440936005600} \over _{6054067458131}}  {\cal K}_2(14)
-{_{59538751957} \over _{90359215793}}       {\cal K}_3(14)  \cr
& \ \ \ \ \ \ \ \ \       \cr
{\cal K}_{15}(14) &\equiv
-{_{140948002575} \over _{553434141824}}   {\cal Q}_{14}
+{_{41019891925} \over _{142682239689}}    {\cal K}_1(14)
-{_{2350269799600} \over _{998775677823}}  {\cal K}_2(14)
+{_{15233364348} \over _{4969033223}}      {\cal K}_3(14)   \cr
& \ \ \ \ \ \ \ \ \       \cr
{\cal K}_{16}(14) &\equiv
-{_{167541965325} \over _{110750985344}}  {\cal Q}_{14}
+{_{16253164725} \over _{9517662803}}     {\cal K}_1(14)
-{_{250019670828} \over _{66623639621}}   {\cal K}_2(14)
+{_{202871157374} \over _{66623639621}}   {\cal K}_3(14)    \cr
& \ \ \ \ \ \ \ \ \       \cr
{\cal K}_{17}(14) &\equiv
-{_{1460008554975} \over _{2154631950208}}  {\cal Q}_{14}
+{_{141634721175} \over _{185163683221}}    {\cal K}_1(14)
-{_{8115082515600} \over _{1296145782547}}  {\cal K}_2(14)
+{_{125668436566} \over _{19345459441}}     {\cal K}_3(14)  \cr
& \ \ \ \ \ \ \ \ \       \cr
{\cal K}_{18}(14) &\equiv
+{_{454756763025} \over _{3325977997952}}   {\cal Q}_{14}
-{_{44115732825} \over _{285826234199}}     {\cal K}_1(14)
+{_{2527648652400} \over _{2000783639393}}  {\cal K}_2(14)
-{_{3254550496} \over _{29862442379}}       {\cal K}_3(14)  \cr
& \ \ \ \ \ \ \ \ \       \cr
{\cal K}_{19}(14) &\equiv
-{_{1971081945} \over _{10746294016}}  {\cal Q}_{14}
+{_{3250632945} \over _{15699663914}}  {\cal K}_1(14)
-{_{13303413960} \over _{7849831957}}  {\cal K}_2(14)
+{_{582923267} \over _{234323342}}     {\cal K}_3(14)  } $$
It is seen that the generators ${\cal Q}_{12}$ and ${\cal Q}_{14}$ play the
role of {\bf a kind of cohomology operators} in the sense ,for instance, that
16 generators ${\cal K}_\alpha(14)$ \ (for $\alpha$ = 4,5, .. 19) will depend
linearly on the first 3 generators ${\cal K}_1(14)$, ${\cal K}_2(14)$ and
${\cal K}_3(14)$ modulo ${\cal Q}_{14}$. It is therefore easy to conclude
that all our 19 generators ${\cal K}_\alpha(14)$ \ (for $\alpha$ = 1,5, .. 19)
are linearly independent due to the fact that there is no a linear
relationship among the generators ${\cal K}_1(14)$, ${\cal K}_2(14)$ and
${\cal K}_3(14)$ modulo ${\cal Q}_{14}$. The similar situation is also true
for degree 12.

\end